# The completed High-Low method for interface state analysis in MOS capacitors

B. D. Rummel[1,a),*], S. Dhar[2], R. J. Kaplar[1]

[1] *Sandia National Laboratories, Albuquerque, New Mexico 87123, USA*

[2] *Department of Physics, Auburn University, Auburn, Alabama 36849, USA*

[a)] *Author to whom correspondence should be addressed: bdrumme@sandia.gov*

Interface state densities, $D_{IT}$, in metal-oxide-semiconductor (MOS) capacitors are rarely reported in the accumulation energy range. It is recognized that the determination of $D_{IT}$ in accumulation is fundamentally obscured by small inaccuracies in the user-defined oxide capacitance, $C_{OX}$. This source of error prevents the High-Low frequency technique from reporting accumulation $D_{IT}$, even for sufficiently fast high-frequency measurements. To resolve this, an electrostatic constraint that is uniquely satisfied by a physically consistent $C_{OX}$ is derived from established theory, thereby completing the High-Low framework. The "completed" framework's theoretical validity is confirmed using simulated capacitance data for an *n*-SiC MOS structure, and the method's frequency limitations are demonstrated. This analytical advancement ensures a physically consistent extraction of $D_{IT}$ near the band edge, overcoming a fundamental limitation in MOS capacitor characterization.

## I. Introduction

The characterization of interface state densities, $D_{IT}$, remains a fundamental requirement for the optimization of metal-oxide-semiconductor (MOS) systems. While an accurate analysis of interface states near the band edges is a universal challenge across all semiconductor materials, the deleterious effects of these defects are particularly restrictive in wide-bandgap (WBG) semiconductors, such as silicon carbide (SiC) and gallium nitride (GaN). These semiconductors are otherwise preferred materials for high-power transistors because of their high-breakdown capabilities and high-temperature resilience.[1-3] Although SiC MOS field-effect transistors (MOSFETs) have achieved commercial success, their on-state performance fails to meet theoretical expectations due in great part to a high $D_{IT}$.[4] As the applied gate bias approaches the threshold condition from the off-state, the subsequently charged interface states compensate as effective gate charge to shift the channel's weak- and strong-inversion modes. In addition, these defects capture free carriers while also behaving as Coulomb scattering centers that reduce channel carrier mobilities, collectively lowering the device's transconductance. These effects limit GaN MOS transistors as well, where similar to SiC, devices are characterized by slow 'turn-on' behavior with higher-than-expected on-resistance.[5, 6]

Interface trap states that hinder an $n$-channel MOSFET are often studied using an $n$-type MOS capacitor. The $n$-type capacitor refers to the conductivity type in the bulk, whereas an $n$-channel MOSFET refers to the conductivity type of the inversion channel. For $p$-channel device development, $p$-type MOS capacitors are used. Figure 1 illustrates the characteristic structures and band diagrams for an $n$-type MOS capacitor biased at the onset of accumulation (or the flat-band condition) and a vertical $n$-channel power MOSFET biased at the onset of strong inversion. Interface states with energies closest to the band edge tend to have greater $D_{IT}$ values, which is conveyed in the acceptor-like $D_{IT}(E)$ plot in Fig. 1. For shallow impurities, if the $n$-type doping concentration in the MOS capacitor equals the $p$-type doping concentration in the MOSFET channel region, then the energy range for accumulation in the MOS capacitor corresponds with the energy range for strong inversion in the MOSFET. Therefore, accumulation interface states for an $n$-type MOS capacitor can correspond to the $D_{IT}$ most relevant to the performance degradation of an $n$-channel MOSFET. Despite this direct relevance, accumulation interface states are rarely reported in MOS capacitor $D_{IT}$ analysis.

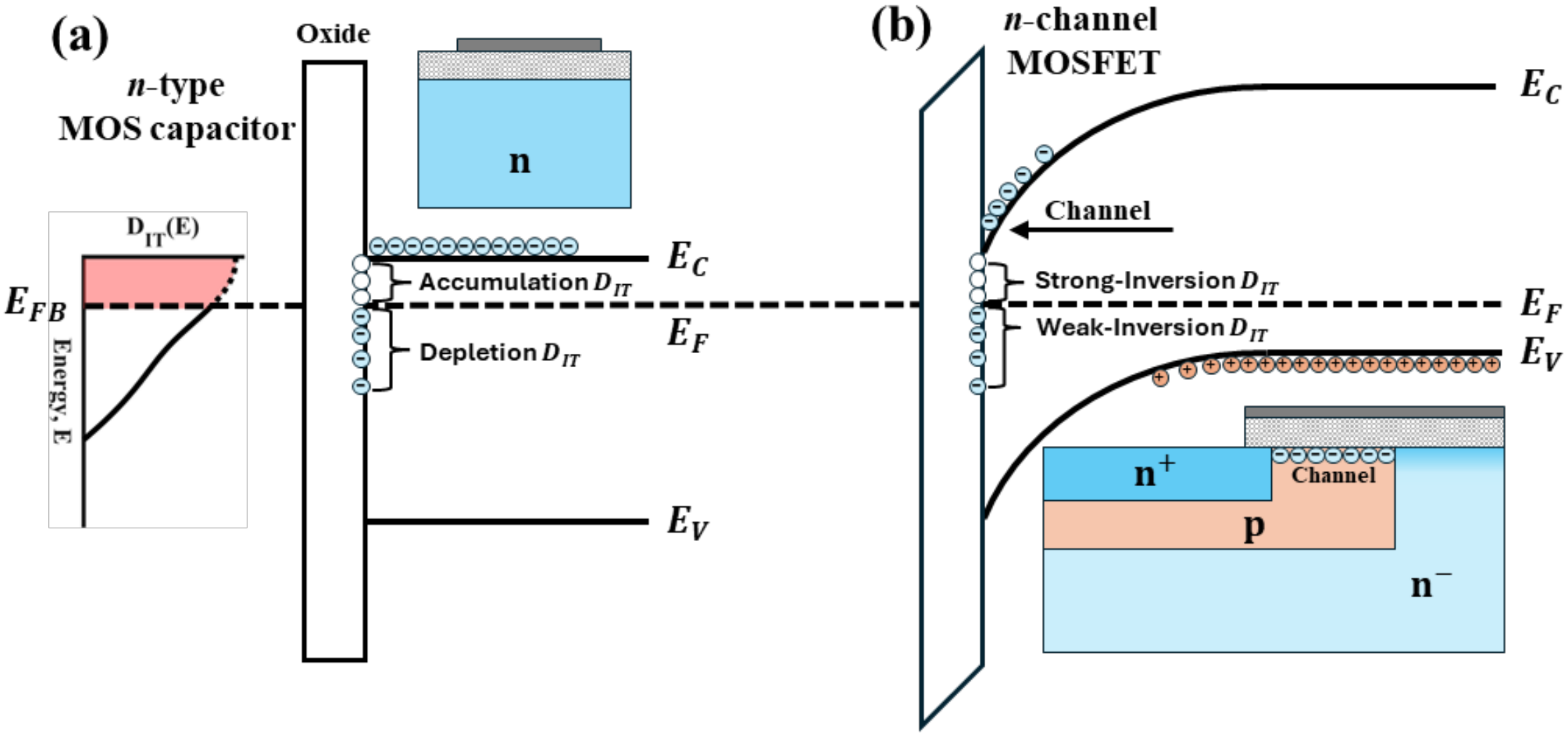

*Figure 1: Band diagrams for (a) an n-type MOS capacitor biased at the flat-band condition and (b) a vertical n-channel MOSFET biased to the onset of strong-inversion. The $D_{IT}(E)$ plot on the left highlights the acceptor-like accumulation traps (red) of an n-type MOS capacitor that align with the traps most relevant to strong inversion in the corresponding n-channel MOSFET.*

The High-Low $C$-$V$ method is a standard technique for extracting $D_{IT}(E)$ from MOS capacitor structures, provided the high-frequency ac signal is sufficiently fast.[7] This work first highlights the known issue that an inaccurate oxide capacitance, $C_{OX}$, is a fundamental source of error that has not previously been resolved. [8-10] Due to this error, the High-Low method, as it is conventionally used, cannot provide an accurate analysis for $D_{IT}$ with energies in accumulation, regardless of how fast the measurement signal is. To resolve this dilemma, an electrostatic constraint is derived from well-established theory to complete the High-Low method. The validity of this constraint is then proven using simulated *n*-SiC data generated from an analytical model, and the completed framework accurately extracts $D_{IT}$ for sufficiently fast high-frequency signals. The frequency limitations of the completed framework are then demonstrated, and analysis recommendations are made to accommodate fast interface states, such as completing the analysis at lower temperatures. While validated here using WBG parameters, the completed framework presented in this work provides a rigorous path for interface state analysis across all relevant MOS material systems.

## II. The High-Low $C$-$V$ Framework

The theory of MOS capacitors is recognized and supported by substantial literature,[8, 10-13] and the following discussions largely draw from these sources. The present analysis addresses phenomena in accumulation and depletion, ignoring inversion, and focuses on *n*-type MOS structures.

### A. The conventional, or "incomplete", framework

The total MOS capacitance, $C_{MOS}$, is described using the small-signal (ac) circuit model, shown in Fig. 2. The oxide capacitance $C_{OX}$ is in series with the semiconductor capacitance, $C_S$, which is itself the parallel sum of the space-charge capacitance, $C_{SC}$, and the interface trap capacitance, $C_{IT}$, i.e., $C_S(\psi_S) = C_{SC}(\psi_S) + C_{IT}(\psi_S)$. Therefore, $C_{MOS}$ is given by

$$C_{MOS} = \left(\frac{1}{C_{OX}} + \frac{1}{C_S}\right)^{-1}. \tag{1}$$

The oxide layer acts as a simple, parallel-plate capacitor such that $C_{OX}$ is independent of the electrostatics within the underlying semiconductor structure. Consequently, $C_{OX}$ is constant with surface potential, $\psi_S$. Then, $C_{SC}$ describes the capacitive response associated with the space-charge (or depletion) region that forms within the semiconductor substrate as a function of $\psi_S$. And lastly, $C_{IT}(\psi_S)$ is defined as

$$C_{IT}(\psi_S) = \left|\frac{dQ_{IT}}{d\psi_S}\right| = qD_{IT}(\psi_S), \tag{2}$$

where $q$ is the fundamental charge constant. A series parasitic impedance, $Z(f_{ac})$, is observed for higher frequency measurements, which may be accounted for in experimental data.[8, 10] A frequency-dependent conductance related to interface states is also part of the small-signal model, functioning in parallel with $C_S$. However, since the High-Low analysis does not consider this conductance, it is omitted in this work.

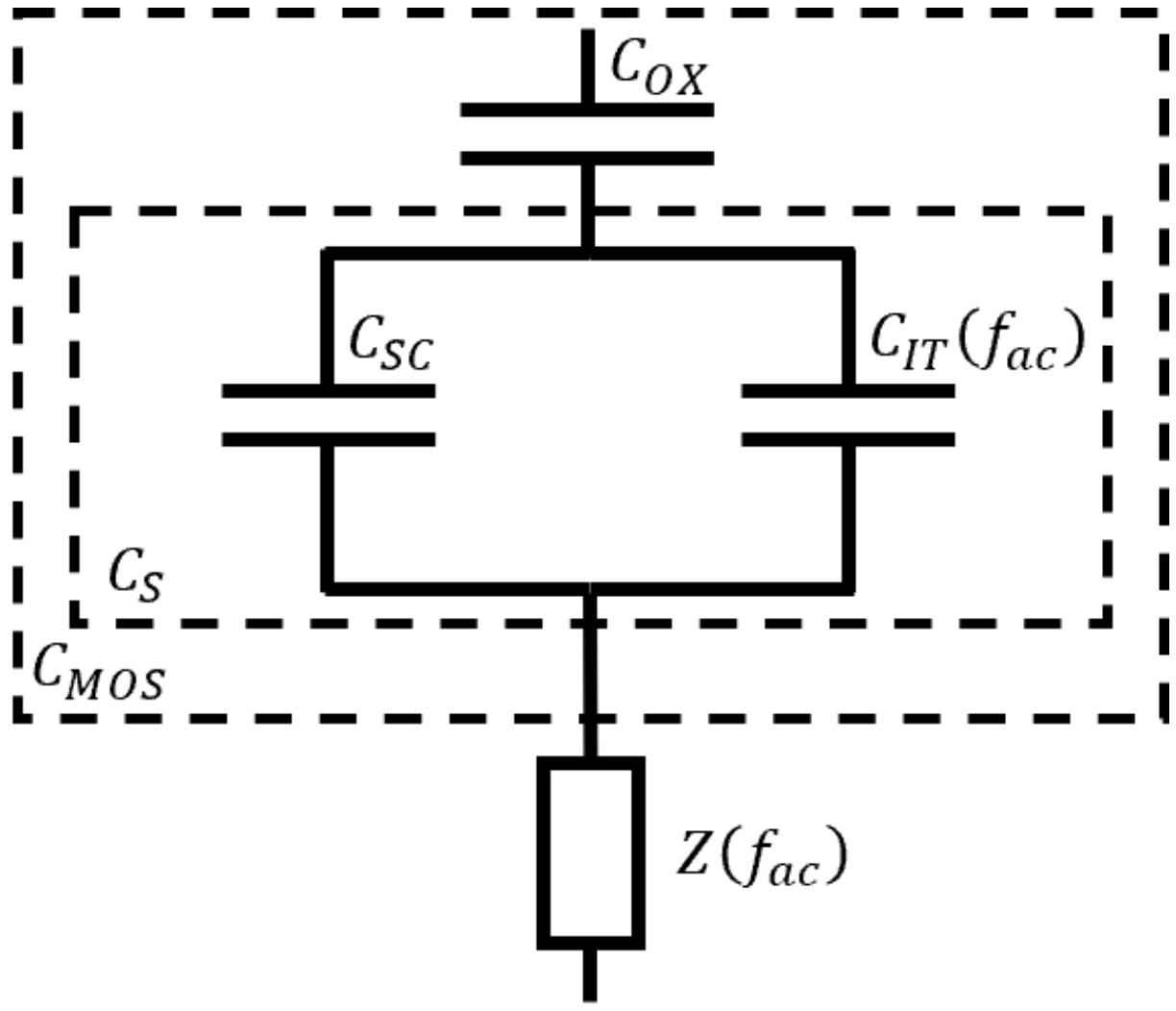

*Figure 2: The small signal model describes $C_{MOS}$ as being the series sum of the oxide capacitance $C_{OX}$ and semiconductor capacitance $C_S$. A series parasitic impedance $Z(f_{ac})$ is observed at higher frequencies.*

Contributions from $C_{IT}$ to the small-signal model vary according to the frequency of the applied ac signal $f_{ac}$. For a low-frequency measurement, interface states with energies over a broad range near the band edge are fast enough to capture and emit charge carriers during the ac oscillation and capacitively contribute to the $C_{MOS}$ measurement. During a higher-frequency measurement, if the ac oscillation is too fast for the interface states to respond, their capacitive contributions are zero. In the High-Low method, a true high-frequency measurement is one where the ac oscillation is fast enough so that all capacitive contributions from interface states are ignored. However, during any measurement, regardless of frequency, the dc-voltage-sweep ramp rate is slow enough such that dc-charged interface states contribute to the $V_G(\psi_S)$ relationship, leading to the $C$-$V$ 'stretch-out' effect. In other words, despite a high-frequency measurement ignoring $C_{IT}(\psi_S)$, $D_{IT}(\psi_S)$ is still encoded in the subsequent $V_G(\psi_S)$ relationship.

The High-Low $C$-$V$ method relies on measurements of both a low-frequency MOS capacitance, $C_{LF}(V_G)$, and a high-frequency MOS capacitance, $C_{HF}(V_G)$, to parameterize an experimental device.[7] A fundamental assumption in utilizing this method is that $C_{HF}$ represents a true high-frequency measurement at least within the measured bounds of $V_G$. Assuming this condition is met, $D_{IT}$ is defined as

$$D_{IT} = \frac{1}{q}\left(C_{S,LF} - C_{S,HF}\right), \tag{3}$$

where $C_{S,LF} = C_{SC} + C_{IT}$ and $C_{S,HF} = C_{SC}$ are the semiconductor capacitances for the low- and high-frequency measurements. Calculation of $C_{S,LF}$ and $C_{S,HF}$ from the experimental $C_{MOS}$ data is done using (1), where a value for $C_{OX}$ is usually obtained by estimation from measurement in accumulation.

Describing $D_{IT}$ for specific trap energies relative to the bandgap first requires that a $\psi_S(V_G)$ relationship be defined. For analyses that utilize a low-frequency device measurement, this may be done using the Berglund equation,[14]

$$\psi_S(V_G) = \int_{V_G^+}^{V_G}\left(1 - \frac{C_{LF}(V_G)}{C_{OX}}\right)dV_G + \psi_S^+, \tag{4}$$

where $V_G^+$ is the maximum gate voltage measured in accumulation and $\psi_S^+$ is the surface potential that corresponds to $V_G^+$. Then, $\psi_S^+$ can be experimentally determined by recognizing that the flat-band voltage, $V_{FB}$, corresponds with the flat-band capacitance, $C_{FB}$, of a 'true' high-frequency measurement, and $\psi_S(V_{FB})$ is equal to 0 V. The flat-band capacitance $C_{FB}$ is defined as

$$C_{FB} = \left(\frac{1}{C_{OX}} + \frac{1}{C_{\text{Debye}}}\right)^{-1}, \tag{5}$$

where $C_{\text{Debye}}$ is the Debye capacitance (see Appendix C). After identifying $V_{FB}$ from the high-frequency curve, $\psi_S^+$ is chosen so that $\psi_S$ is 0 V at $V_G = V_{FB}$. Finally, the calibrated $\psi_S$ is linearly translated to the trap energy, $E_T$, described with respect to the relevant bulk band edge (e.g., the conduction band edge, $E_C$, for an *n*-type semiconductor). The $E_C - E_T(\psi_S)$ relationship is defined as

$$E_C - E_T(\psi_S) = \left(\frac{E_g}{2} - \psi_f\right) - \psi_S, \tag{6}$$

where $E_g$ is the semiconductor bandgap and $\psi_f$ is the bulk Fermi level.

## b. Completing the High-Low framework

It is well known that small variations in the user-approximated $C_{\text{OX}}$ greatly affect the calculation of $D_{\text{IT}}$ in accumulation using any capacitance-based method.[8-10] This is plainly evident in the relationship for $C_{\text{S}}$, rewritten from (1),

$$C_S = \frac{C_{\text{OX}} C_{MOS}}{C_{\text{OX}} - C_{MOS}}. \tag{7}$$

Regardless of the capacitive contribution from $C_{IT}$ to $C_S$, when $C_{MOS}$ is measured in accumulation so that it is nearer to $C_{\text{OX}}$, the determination of $C_S$ is particularly sensitive to the choice for $C_{\text{OX}}$ because $C_{\text{OX}} - C_{MOS}$ approaches zero while the numerator does not. Consequently, slight deviations in $C_{\text{OX}}$ from the real value have a considerable impact on the error in calculating $D_{\text{IT}}$ for trap energies very close to the band edge (i.e., in accumulation) and, to a lesser degree, for energies deeper in the bandgap. Furthermore, the mapping from $V_G$ to $E_C - E_T$ is affected by the $C_{\text{OX}}$ dependence in (4) and (5). Therefore, poor approximation of $C_{\text{OX}}$ results in uncertainty when calculating both the $D_{IT}$ near the band edge and the trap energy level when using $C$-$V$ methods.

A simplistic method to approximate $C_{\text{OX}}$ is by precisely measuring the dielectric thickness, $t_{OX}$, and then calculating $C_{\text{OX}} = \epsilon_{\text{OX}}\epsilon_0/t_{OX}$, where $\epsilon_0$ is the permittivity of free space and $\epsilon_{\text{OX}}$ is the relative permittivity of the oxide. However, to be applicable for $C$-$V$ analysis, $\epsilon_{OX}$ must be known with great precision, which necessitates reverse calculation starting with an accurate $C_{\text{OX}}$. Another common though misguided practice is to estimate $C_{\text{OX}}$ as the maximum low-frequency capacitance measured far in accumulation, i.e., $C_{\text{OX}} = C_{LF}(V_{\text{G}}^{+})$. Although $C_{LF}(V_G^{+})$ can approach $C_{OX}$ in the presence of extreme trap densities, imposing a strict equivalence between the two is an invalid analytical constraint. This assumption pre-determines the outcome by implying an infinitely large $C_S(V_G^{+})$, which forces the extracted $D_{IT}$ to appear infinite near the band edge. Such an approach risks reporting a high $D_{IT}$ in accumulation as an artifact of the mathematical assumption rather than a physical reality of the interface. An approximation that $C_{\text{OX}} = f_{\text{OX}} C_{LF}(V_G^{+})$, where $f_{OX}$ is a correction factor, may improve the user-estimated $C_{OX}$, but there has not previously been a method to indicate a correct value for $f_{\text{OX}}$. These errors become more pronounced (or $f_{OX}$ gets larger) with increasing $C_{\text{OX}}$, as $C_{\text{MOS}}$ saturates in accumulation sooner as a consequence of Fermi-Dirac statistics and quantum confinement effects.[15] Without a means to determine a physically consistent $C_{\text{OX}}$, the system of equations to conduct $C$-$V$ analysis is underdetermined.

To complete the High-Low framework, an electrostatic constraint is derived using fundamental relationships from established theory. During device operation, the externally applied $V_G$ controls the internal system parameter $\psi_S$. According to Kirchhoff's voltage law, the gate voltage $V_G$ is the sum of voltage drops across the oxide layer, $V_{OX}$, the semiconductor, $V_S$, and the built-in potential arising from the metal-semiconductor work function difference, $\phi_{MS}$. The gate voltage $V_G$ is therefore related to $\psi_S$ by the following:

$$V_G(\psi_S) = \phi_{MS} + V_{OX}(\psi_S) + V_S(\psi_S), \tag{8}$$

which, when considering acceptor-like interface traps, is expanded as

$$V_G(\psi_S) = \left[\phi_{MS} - \frac{Q_{OX}}{C_{OX}} + \frac{q}{C_{OX}}\int_{-\infty}^{\psi_S} D_{IT}(\psi_S)d\psi_S\right] + \psi_S - \frac{Q_S(\psi_S)}{C_{OX}}. \tag{9}$$

The bulk dielectric charge, $Q_{OX}$, describes the static charges in the oxide (e.g., fixed, mobile, trapped) treated as an effective sheet charge, which remains constant with respect to $\psi_S$. A brief discussion on the well-established derivation of (9) can be found in Appendix A. At the flat-band condition, or $\psi_S = 0$, the unbracketed terms evaluate to zero and the bracketed terms describes $V_{FB}$ as

$$V_{FB} = \phi_{MS} - \frac{Q_{OX}}{C_{OX}} + \frac{q}{C_{OX}}\int_{-\infty}^{0} D_{IT}(\psi_S)d\psi_S\,. \tag{10}$$

Notice that $V_{FB}$ does not consider voltage drop due to interface states that exchange charge in accumulation, i.e., $D_{IT}(\psi_S > 0)$. Substituting (10) into (9) gives:

$$V_G(\psi_S) = V_{FB} + \frac{q}{C_{OX}}\int_{0}^{\psi_S} D_{IT}(\psi_S)d\psi_S + \psi_S - \frac{Q_S(\psi_S)}{C_{OX}}. \tag{11}$$

By evaluating (11) at the maximum surface potential $\psi_S^+$, a relationship is established between the device electrostatics and the maximum applied $V_G^+$ in accumulation,

$$V_G^+ = V_{FB} + \frac{q}{C_{OX}}\int_{0}^{\psi_S^+} D_{IT}(\psi_S)d\psi_S + \psi_S^+ - \frac{Q_S^+}{C_{OX}}, \tag{12}$$

where $Q_S^+$ is $Q_S(\psi_S)$ solved at $\psi_S^+$. The term $\psi_S^+$ is also the integration constant in (4). Finally, defining $\Delta V_{IT}^+ = \frac{q}{C_{OX}}\int_0^{\psi_S^+} D_{IT}(\psi_S)d\psi_S$ and rearranging (12) gives

$$\Delta V_{IT}^{+} = (V_G^{+} - V_{FB}) - \left(\psi_S^{+} - \frac{Q_S^{+}}{C_{OX}}\right). \tag{13}$$

Equation (13) states that the voltage drop due to accumulation interface charge, $\Delta V_{IT}^{+}$, is equal to the flat-band-referenced voltage drop across the MOS structure, $V_G^{+} - V_{FB}$, minus both the semiconductor voltage drop, $\psi_S^{+}$, and the oxide voltage drop from accumulated semiconductor charge, $-\frac{Q_S^{+}}{C_{OX}}$. Fundamentally, this relationship enforces the conservation of energy across the structure via Kirchhoff's voltage law, establishing a strict electrostatic boundary. This constraint is derived from relationships that are inherent to MOS operation, but it is not considered within the conventional High-Low method. Therefore, (13) provides the final constraint necessary to complete the High-Low framework without relying on physically arbitrary assumptions. A unique and physically consistent $C_{\mathrm{OX}}$ may be identified through a simple guess-and-check algorithm, illustrated in Fig. 3. The flowchart for this algorithm shows that the conventional High-Low analysis is a subset of the completed High-Low analysis. The determination of $C_{\mathrm{OX}}$ is demonstrated in the following section and visualized graphically.

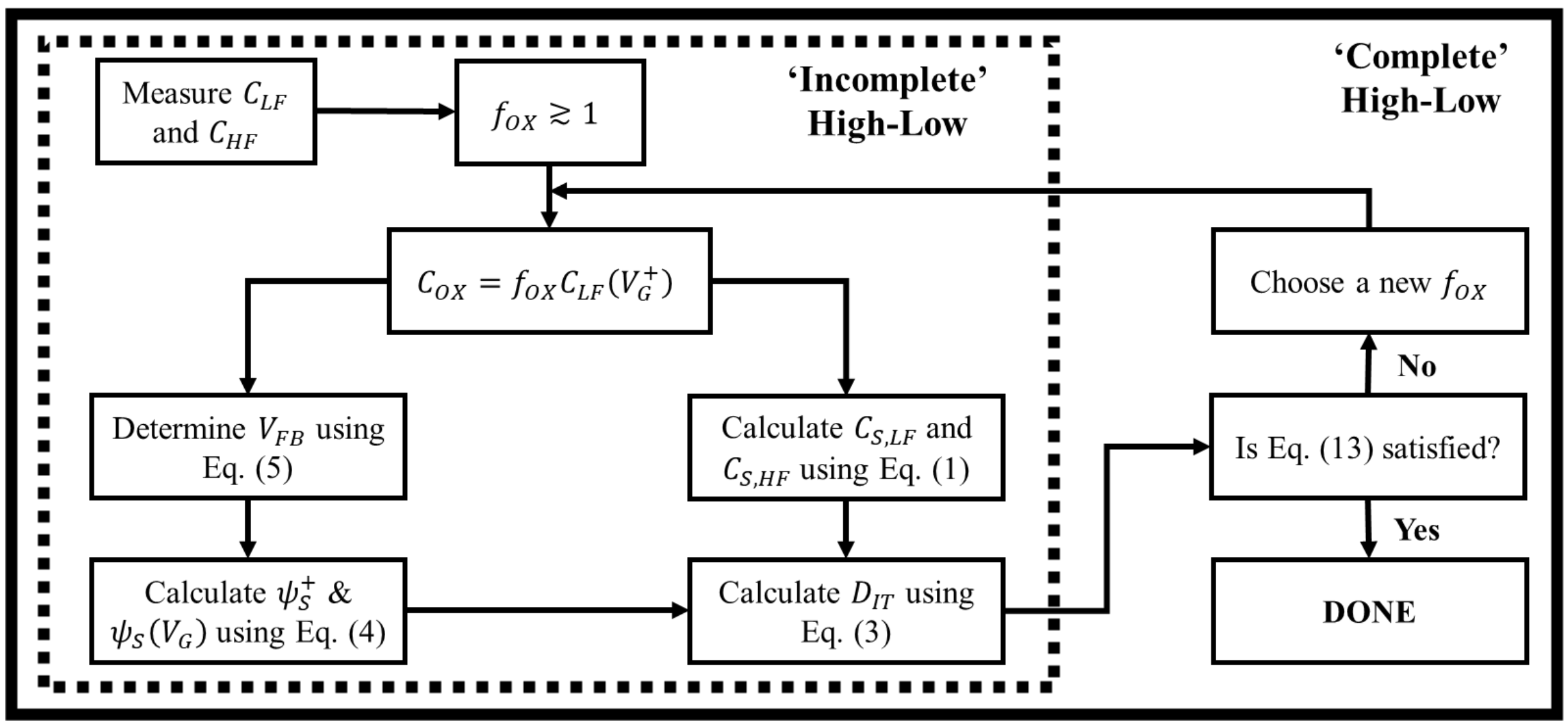

*Figure 3: A flowchart showing the conventional, or "incomplete", High-Low framework, as it is conventionally used, as a subset of the "complete" High-Low framework. The symbol $\gtrsim$ is used to describe the correction factor $f_{OX}$ as being approximately equal to though strictly greater than 1.*

To calculate the right-hand side of (13), the $Q_S(\psi_S)$ relationship must be known. Neglecting quantum confinement effects, an analytical expression for $Q_S(\psi_S)$ can be described by solving Poisson's equation for the various operating modes of the MOS structure. When a device is operated such that the surface Fermi level is within $3kT$ of the band edge or is outside of the bandgap, the Maxwell-Boltzmann approximation is invalid and Fermi-Dirac statistics are used. The semiconductor charge $Q_S(\psi_S)$ is then defined for the degenerate case as

$$Q_S(\psi_S) = -\mathrm{Sign}(\psi_S)\frac{kT}{q}\frac{\epsilon_S\epsilon_0}{L_D}$$

$$\times\sqrt{\exp\left(\frac{q}{kT}\frac{E_g}{2}\right)\left[\mathcal{F}_{3/2}(\eta_S(\psi_S)) - \mathcal{F}_{3/2}(\eta_0)\right] - \exp\left(\frac{q\psi_f}{kT}\right)[\eta_S(\psi_S) - \eta_0]}, \tag{14}$$

where $k$ is the Boltzmann constant, $T$ is temperature, $\epsilon_S$ is the relative permittivity of the semiconductor, $L_D$ is the intrinsic Debye length, and $\psi_f$ is the bulk Fermi potential. The Fermi-Dirac integral of order $j$, $\mathcal{F}_j(\eta_S)$, is either solved using numerical methods or an analytical approximation.[14] A discussion on calculating $\mathcal{F}_j(\eta_S)$ is found in Appendix B. The normalized surface Fermi level position referencing the conduction band edge, $\eta_S$, is given by

$$\eta_S(\psi_S) = -\frac{q}{kT}\left(\left(\frac{E_g}{2} - \psi_f\right) - \psi_S\right), \tag{15}$$

and $\eta_0 = \eta_S(0)$ is the normalized bulk Fermi level with respect to the conduction band edge.

When carrier confinement at the interface necessitates quantum confinement considerations, $Q_S^+$ must be calculated via a self-consistent solution of the Schrödinger and Poisson equations.[15-17] In MOS structures, particularly under the strong electric fields required for moderate or strong accumulation, quantum confinement alters the spatial distribution of the carriers near the interface. Consequently, calculating $Q_S^+$ using the classical approximation in (14) overestimates the actual semiconductor charge. While (13) remains a fundamental electrostatic requirement regardless of the specific charge distribution, the accuracy of the analysis is directly tied to the fidelity of the $Q_S$ model. Thus, correctly incorporating quantum confinement effects, along with higher-order physical refinements such as vertical doping gradients when present, yields a more physically precise result. In the following section, quantum confinement considerations are omitted, and $Q_S$ is defined using (14) strictly for the purpose of demonstrating the completed framework.

Beyond vertical quantum confinement, a further consideration is the presence of lateral non-uniformities due to randomly distributed interface charges. In the depletion regime, the surface potential is highly sensitive to these local charge variations.[18-21] The measured capacitance reflects a spatial average of these variations, introducing a stretch-out effect that one-dimensional analysis misattributes to interface states, thereby overestimating $D_{IT}$ in this regime.[22, 23] Because the conductance method explicitly accounts for this statistical variance,[12] it remains the standard technique for depletion analysis. However, under strong accumulation, the surface potential becomes weakly dependent on gate voltage, forcing localized regions of the device to be biased to approximately the same surface potential. This effective pinning suppresses lateral variations and preserves the macroscopic validity of the one-dimensional electrostatic description in (13). Consequently, the completed High-Low method exploits this uniformity, providing a complementary technique uniquely suited to extract $D_{IT}$ in the accumulation region.

## III. Numerical Verification of the Completed High-Low Method

### A. Definition of the simulated MOS system

A self-consistent analysis that incorporates (13) is demonstrated using simulated MOS capacitor data so that the extracted parameters can be verified with the known input parameters. The analytical model emulates an *n*-type 4H-SiC MOS capacitor at room temperature where the uniform bulk donor impurity concentration, $N_D$, is $1 \times 10^{16}$ cm$^{-3}$, $Q_{OX}$ and $\phi_{MS}$ are both zero, and $C_{OX}$ is 100 nF/cm$^2$ to represent a 35-nm $SiO_2$ gate oxide. An analytical expression for the degenerate $C_{SC}$ as well as the material properties used in this simulation are described in Appendices C and D, respectively. The input $D_{IT}$ distribution and energy-dependent interface trap time constant distribution, $\tau_{IT}$, are both shown in Fig. 4. The values are chosen to agree with the distributions reported by Yoshioka et al., which were determined experimentally between $E_C - E_T = 0.16$ eV and 0.47 eV using the conductance technique.[18] For $D_{IT}$ outside of the reported energy range, a fitted distribution is assumed based on a two-Gaussian model yielding $D_{IT}$ that increases as $E_T$ moves closer to the band edge and decreases for $E_T$ further in the bandgap. As a general approximation, $\tau_{IT}$ has an Arrhenius relationship with energy, which is reflected in the regions outside of the reported energy range.

The capacitive contribution from interface traps to the simulated device capacitance is calculated by

$$C_{IT}(\psi_S) = qD_{\mathrm{IT}}(\psi_{\mathrm{S}})\left(1 - \exp\left(\frac{-1}{f_{ac}\tau_{IT}(\psi_S)}\right)\right), \tag{16}$$

where the exponential term acts as an ‘on/off’ function that imparts the characteristic effects of the input ac frequency $f_{ac}$. Similarly, the interface-state charge contribution to the $V_G(\psi_S)$ relationship is calculated by

$$Q_{IT}(\psi_S) = -q \int_{-\infty}^{\psi_S} D_{\mathrm{IT}}(\psi_{\mathrm{S}}) \left(1 - \exp\left(\frac{-\tau_{dc}}{\tau_{IT}(\psi_S)}\right)\right) d\psi_{\mathrm{S}}\,, \qquad (17)$$

where $\tau_{dc}$ is the characteristic residence time for the dc sweep at each gate voltage step. In a real measurement, $\tau_{dc}$ depends on the dc sweep rate and the $V_G$ step size, and in this simulation, $\tau_{dc}$ is 1 s. According to the simulation results, the flat-band energy $E_{FB}$ is about 0.2 eV below $E_C$ and the flat-band voltage is 1.46 V.

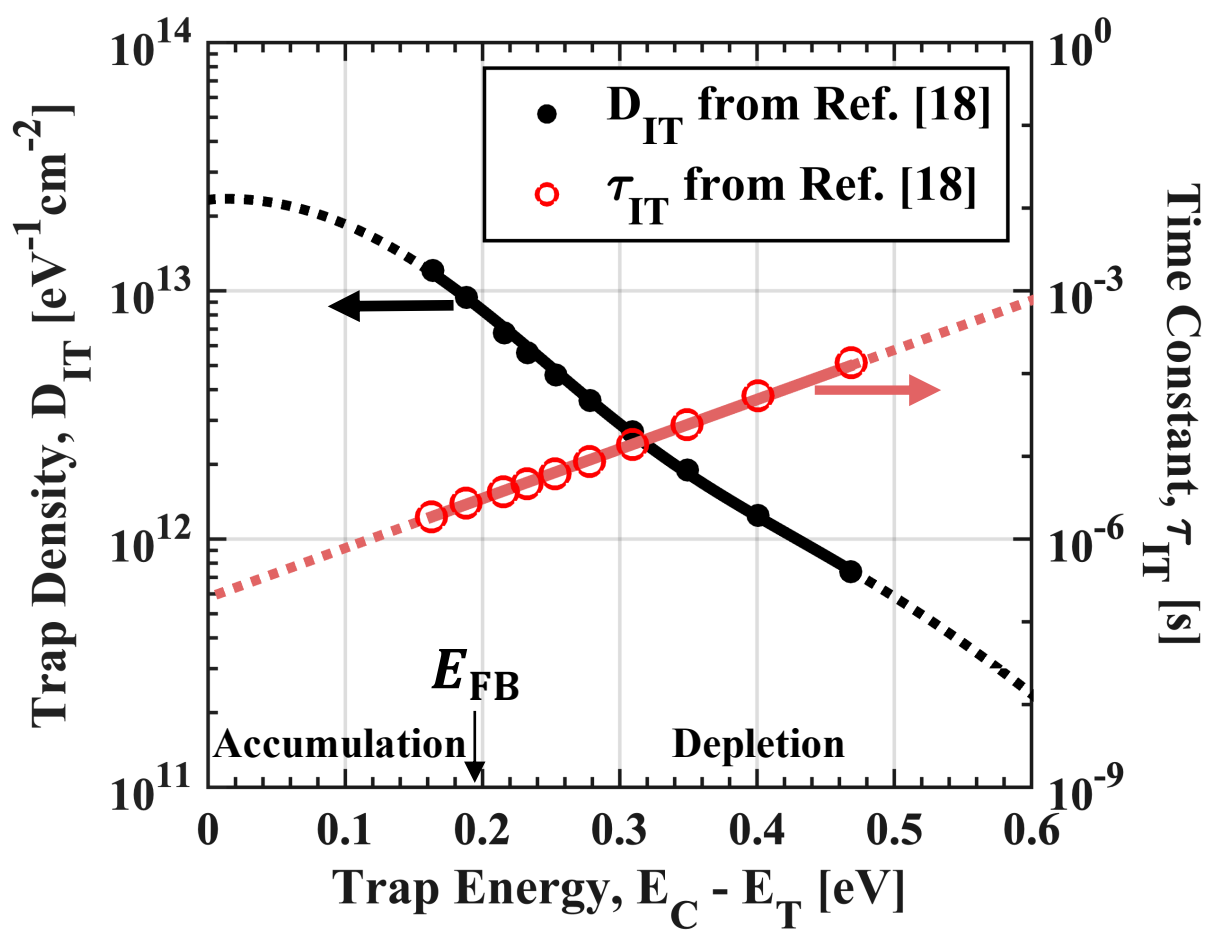

*Figure 4: The analytical model's input $D_{IT}$ and $\tau_{IT}$ distributions are chosen to agree with values reported by Yoshioka et al.[18] The dotted lines are extrapolations beyond the reported ranges. Energies to the left and right of $E_{FB}$ are in accumulation and depletion, respectively.*

To emulate a real device measurement, a finely spaced range of $\psi_S$ broadly spans the operating modes of strong accumulation, depletion, and deep depletion, and the corresponding $C_{SC}(\psi_S)$, $C_{MOS}(\psi_S)$ and $V_G(\psi_S)$ are calculated using (C6), (1) and (9), respectively. Consistent with the one-dimensional framework established in Sec. II, this simulation explicitly omits lateral non-uniformities. Then, $C_{MOS}$ values are tabulated alongside $V_G$ values from $-10$ V to 10 V with a step size of 100 mV. The simulated $C$-$V$ curves for $f_{ac} = 1$ Hz, 100 kHz, 1 MHz, 10 MHz, and 100 MHz are shown in Fig. 5. A simulated 1-THz signal is also included to represent a “true” high-frequency measurement, which is fast enough to ignore the capacitive contribution from all interface states within the $V_G$ limits of the generated data. Notice that the 1-THz and 100-MHz curves nearly overlap. The inset in Fig. 5 shows that the $C_{MOS}$ curves near $V_G^+ = 10$ V do not exceed 99 $\mathrm{nF/cm^2}$, and for moderately higher $V_G^+$ (not shown), $C_{MOS}$ appears to stop increasing about 1% lower than the actual value of $C_{OX}$. Furthermore, the slight frequency dispersion observed between the $C_{MOS}$ curves in the accumulation regime is a direct consequence of weak Fermi level pinning, an effect that becomes pronounced for high $D_{IT}$ values in accumulation.[24] When applying the completed High-Low framework to experimental measurements, it must be noted that improper parasitic correction routines can artificially distort these accumulation characteristics and invalidate the extracted $D_{IT}$.[24]

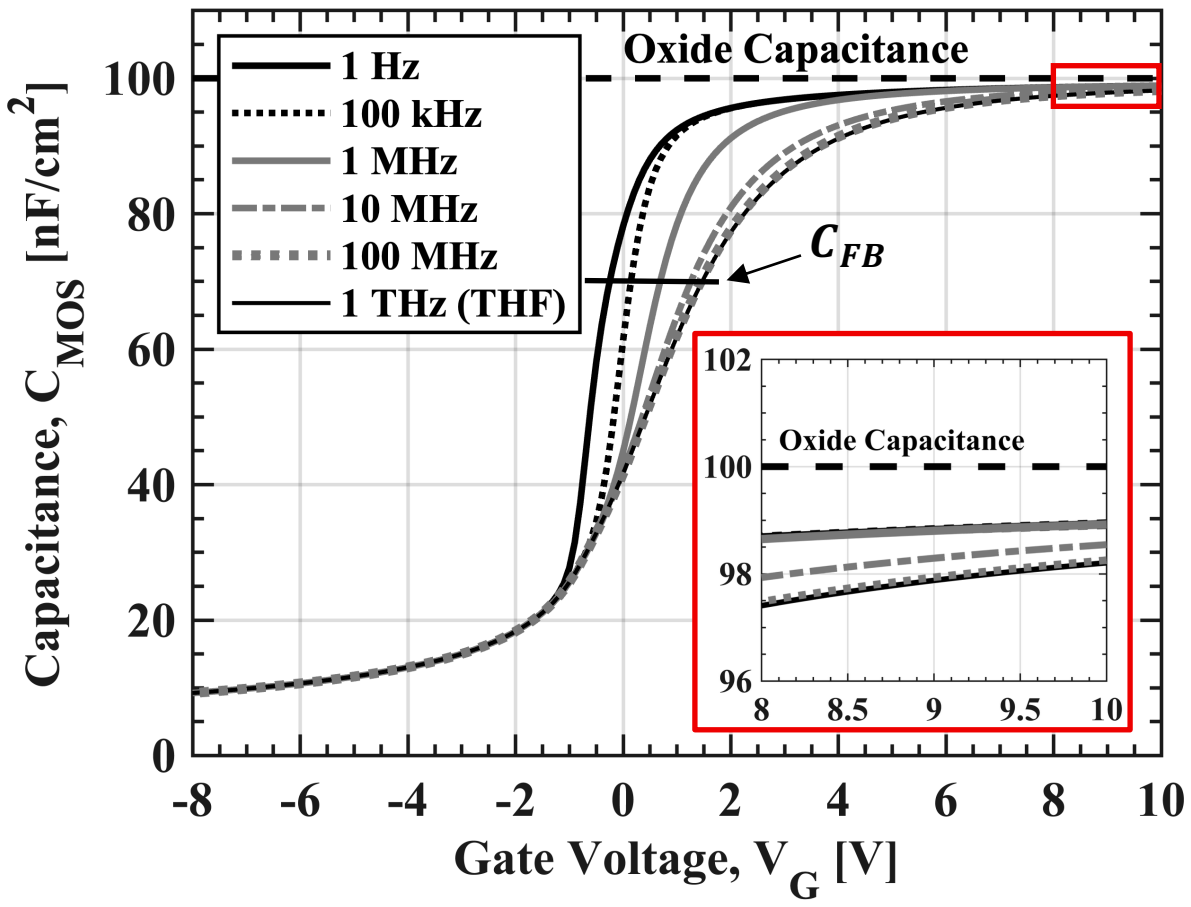

*Figure 5: Frequency-dependent C-V curves are modeled using the $D_{IT}$ and $\tau_{IT}$ distributions shown in Fig. 4. The true high-frequency (THF) signal uses a 1-THz input. The $C_{FB}$ at 70.0 $\mathrm{nF/cm^2}$ corresponds with a $V_{FB}$ of 1.46 V for the 1-THz curve. The inset shows $C_{MOS}$ values near $V_G^+$.*

## B. Extraction of interface state densities

The conventional (or incomplete) High-Low analysis is demonstrated using the true high-frequency (1 THz) and the 1-Hz low-frequency curves. The analysis is done for $C_{OX}$ values ranging from 99.0 $\mathrm{nF/cm^2}$ to 102.0 $\mathrm{nF/cm^2}$, and the corresponding $D_{IT}$ solutions are shown in Fig. 6(a). For the typical assumption that $C_{OX} \approx C_{LF}(V_G^+)$, i.e., 99.0 $\mathrm{nF/cm^2}$, the conventional High-Low analysis reports a seemingly infinite $D_{IT}$ for trap energies close to the band edge. As $C_{OX}$ increases, $D_{IT}$ near the band edge decreases, but only when the $C_{OX}$ guess value matches the input value, i.e., 100.0 $\mathrm{nF/cm^2}$, does the extracted $D_{IT}$ match the input $D_{IT}$. *Overall, these results demonstrate the most fundamental limitation of the conventional High-Low analysis, in that even a true high-frequency measurement is unable to reliably recover $D_{IT}$ in accumulation without a physically consistent $C_{OX}$.*

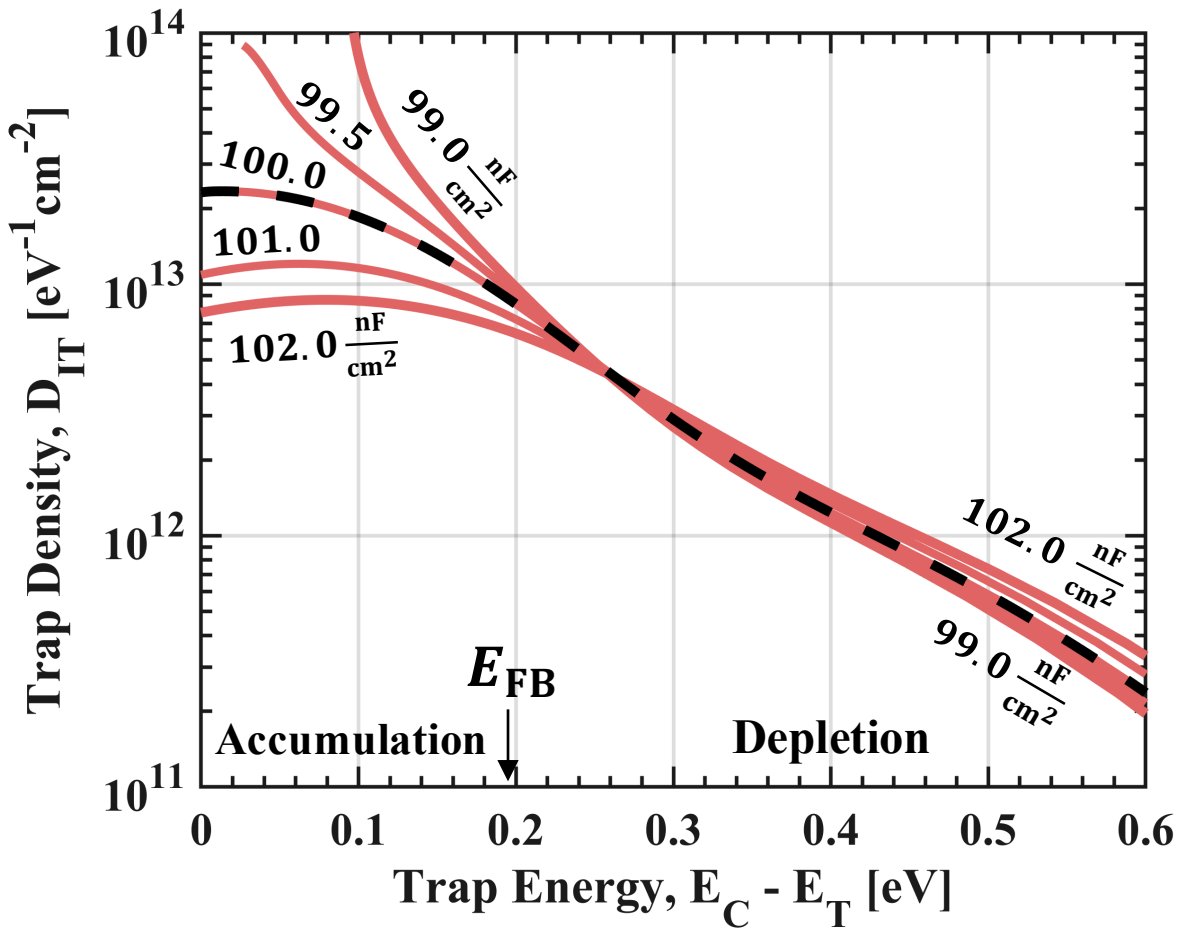

*Figure 6: "Incomplete" $D_{IT}$ solutions (red) are solved using the simulated true high-frequency and 1-Hz low-frequency curves for $C_{OX}$ between 99.0 $nF/cm^2$ and 102.0 $nF/cm^2$. The input $D_{IT}$ (black dash) is only recovered when the input $C_{OX}$, i.e., 100.0 $nF/cm^2$, is used.*

The newly derived electrostatic constraint resolves this limitation, and its application is well visualized using the graphical technique illustrated in Fig. 7. First, $\Delta V_{IT}^{+}$ may be calculated for any value of $C_{OX}$ by numerically integrating the corresponding $\frac{q}{C_{OX}} D_{IT}(\psi_S)$ distribution between 0 and $\psi_S^{+}$. A set of numerically calculated $\Delta V_{IT}^{+}$ is then plotted (solid line) for a range of $C_{OX}$ values. Next, at each $C_{OX}$ value over the same range, the subsequent $V_{FB}$ and $\psi_S^{+}$ values are used to analytically calculate $\Delta V_{IT}^{+}$ value using (13), where $V_G^{+}$ is 10 V and $Q_S^{+}$ is solved for using (14). This second set of $\Delta V_{IT}^{+}$ values is plotted (dashed line) alongside the first. The electrostatic constraint is satisfied at the intersection of these two curves, recovering the input $C_{OX}$ and, consequently, the correct $D_{IT}$ solution shown in Fig. 6. *Thus, the completed High-Low framework enables a physically consistent extraction of $C_{OX}$ and measures $D_{IT}$ into accumulation when using a true high-frequency measurement.*

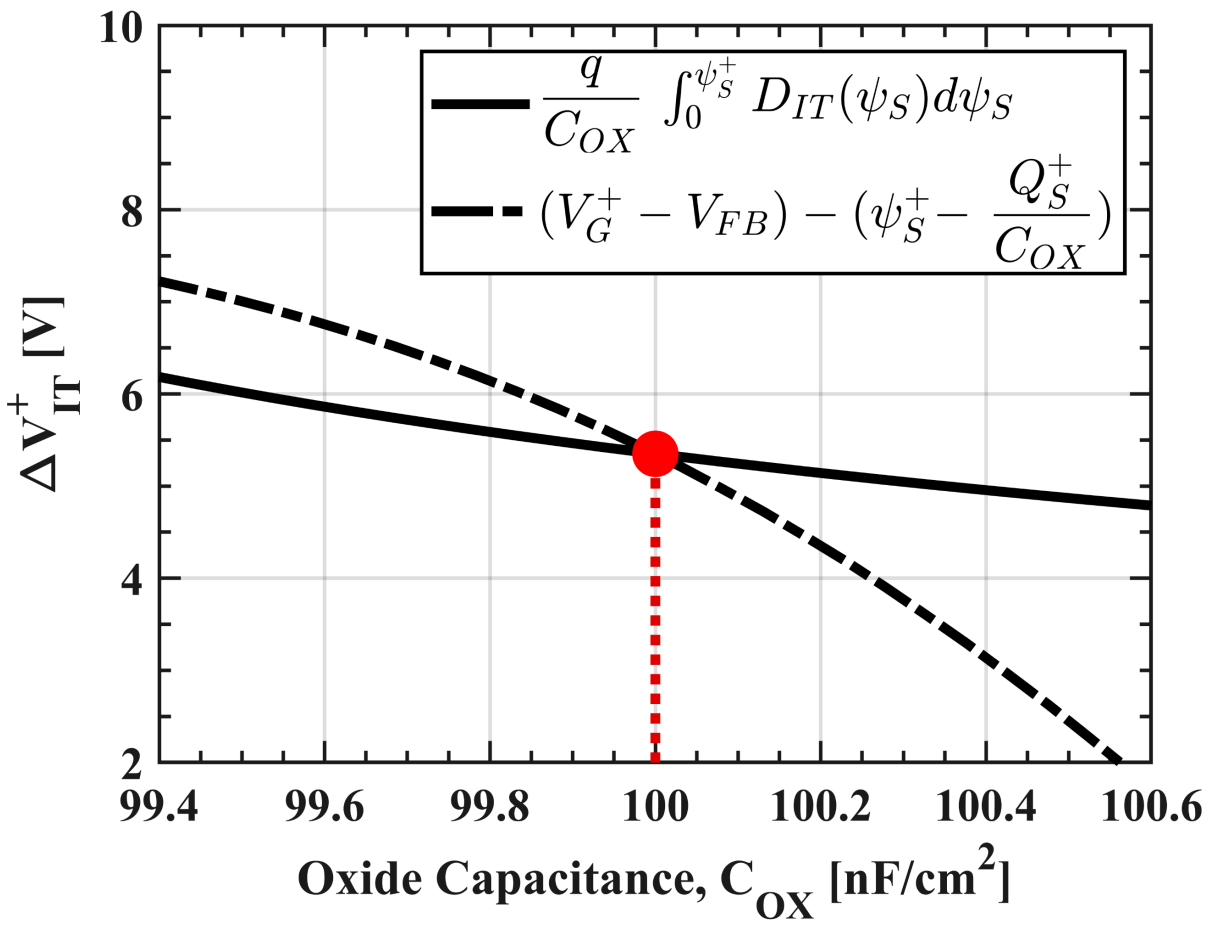

*Figure 7: $\Delta V_{IT}^{+}$ is solved both numerically from $D_{IT}$ (solid) and analytically from* (13) *(dashed) over a range of $C_{OX}$. Their intersection at $C_{OX} = 100.0\, nF/cm^2$ describes when the boundary constraint is satisfied, and the corresponding $D_{IT}$ distribution in Fig. 6 matches the input $D_{IT}$ distribution.*

## IV. Accommodating Fast Traps using the Completed High-Low Method

While the electrostatic constraint provides a completed picture of the High-Low framework, the analysis still requires that the high-frequency measurement be representative of a true high-frequency response. To demonstrate this, analyses are conducted using the 100-kHz, 1-MHz, 10-MHz, and 100-MHz high-frequency data and the 1-Hz low-frequency data. The subsequent $D_{IT}$ results are shown in Fig. 8. Application of the completed High-Low technique using a 100-kHz or 1-MHz signal fails to adequately recover the input $D_{IT}$ throughout the entire energy range. An insufficiently fast signal can still detect $D_{IT}$ for slower traps that do not capacitively contribute to the high-frequency measurement. However, a suitably accurate $V_{FB}$ will not be recovered, which impacts the calculations of $\psi_S$ and, consequently, $E_C - E_T$. Therefore, the subsequent $D_{IT}$ distribution is negatively shifted on the energy axis and is under reported near the band edge. On the other hand, the fast signals for the 10-MHz and 100-MHz measurements enable a satisfactory determination of $V_{FB}$ and their $D_{IT}$ results are nearly converged with the input $D_{IT}$ near the conduction band edge. Also, the extracted $C_{OX}$ values for these frequencies recover the input $C_{OX}$ within 0.1%.

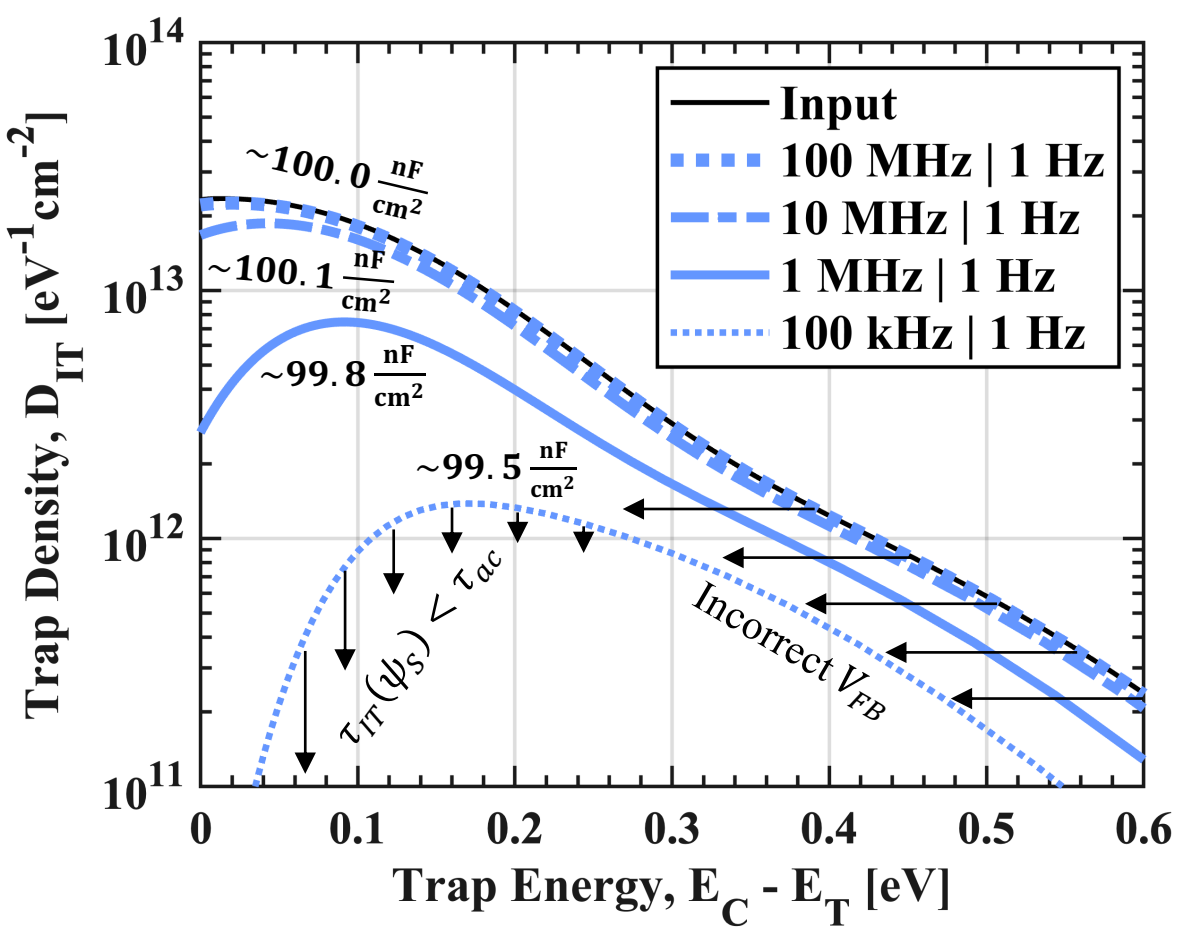

*Figure 8: "Completed" High-Low $D_{IT}$ solutions using the simulated high-frequency measurements and the 1-Hz low-frequency measurement with their extracted $C_{OX}$ values overlaid. For insufficiently fast high-frequency measurements, $D_{IT}$ is negatively shifted on the energy axis due to an incorrect $V_{FB}$. Also, $D_{IT}$ drops near the band edge due to the $\tau_{IT}(\psi_S)$ being smaller than the high-frequency signal time constant $\tau_{ac}$ (or $1/f_{ac}$).*

However, the errors in $D_{IT}$ for each of the high-frequency signals are contingent on the values of $\tau_{IT}(\psi_S)$. Here, the $\tau_{IT}(\psi_S)$ distribution is assumed from an Arrhenius fit of the values reported by Yoshioka et al.,[18] resulting in signals as fast as 10 MHz being sufficient to extract $D_{IT}$ within a reasonable margin of error. However, the time response for interface traps is inherent to the trap species present, and, according to extensive literature, the reported values for $\tau_{IT}(\psi_S)$ can vary substantially within and across different characterization techniques and WBG MOS systems.[18-21] This is illustrated in Fig. 9 for 4H-SiC. In a separate report, Yoshioka et al. describe very fast trap states that are not detectable using a 10-MHz signal at room temperature, which they attribute to a nitrogen interstitial at the SiC surface as a result of nitric oxide annealing.[21] Therefore, only if the high frequency is sufficiently fast relative to the frequency response of the traps, the completed High-Low method can be used to accurately determine $D_{IT}$ in accumulation. It should also be noted that this $D_{IT}$ would not exclude capacitive contributions from slow border traps in the near-interfacial region, as they can also exchange charge during the low-frequency measurement.[25, 26]

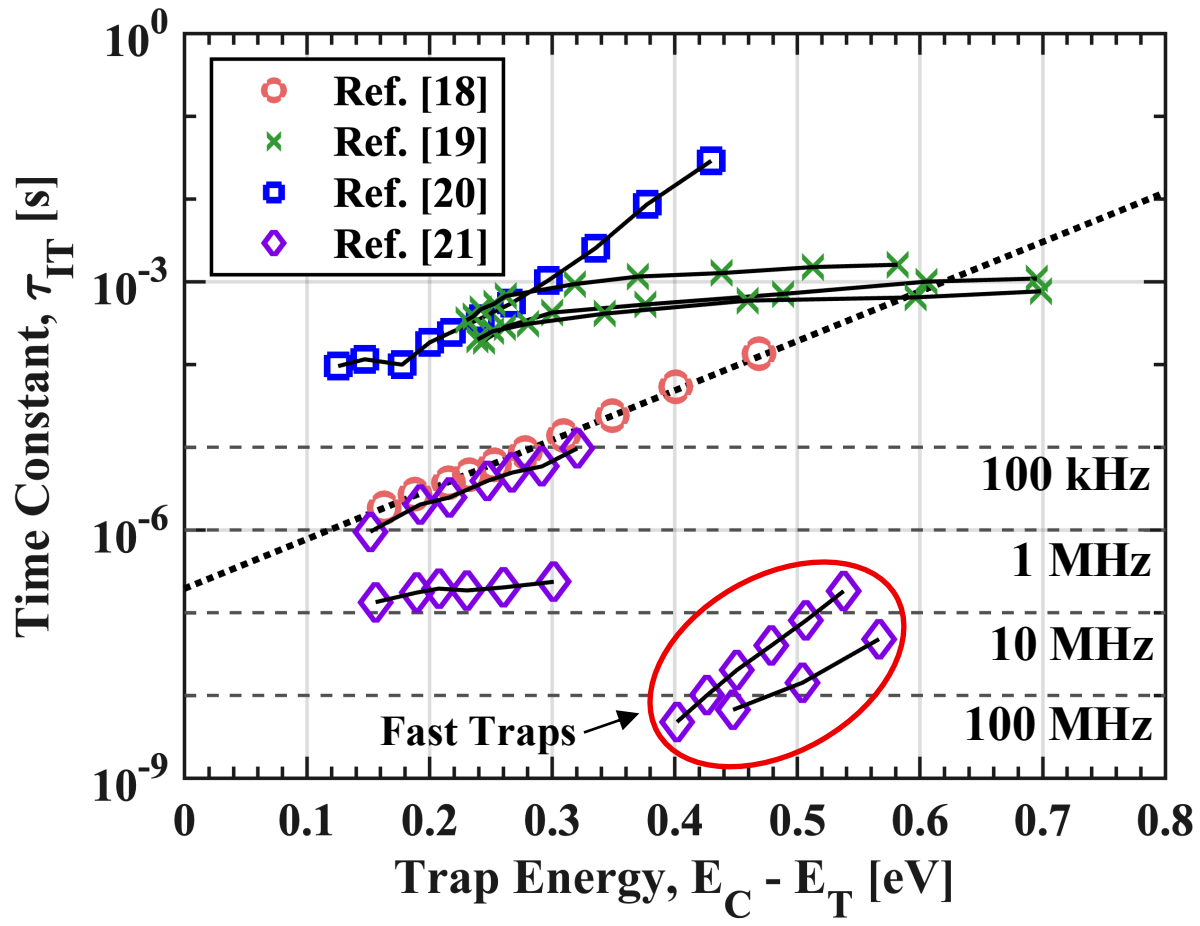

*Figure 9: The range of reported values for $\tau_{IT}(\psi_S)$ at room temperature in $SiO_2$/n-SiC MOS structures tend to increase exponentially with energy. The circled data refers to fast traps created by interfacial nitridation. The dashed line is the Arrhenius fit $\tau_{IT}$ distribution used in this work.*

For MOS systems containing fast interface traps, a 1-MHz measurement is insufficient at room temperature, and the completed High-Low framework does not yield accurate results. In the absence of routine measurements exceeding 10 MHz, a quasistatic $C$-$V$ method (i.e., the $C$-$\psi_S$ method) has been proposed.[18] However, it is essential the correct form of $C_{SC}$ that considers Fermi-Dirac statistics and quantum confinement effects is used as appropriate, and even still there are other unresolved issues in reporting $D_{IT}$ in accumulation.[27] To extend the range of accessible trap energies, performing measurements at lower temperatures can increase trap response times.[21] In yet another report by Yoshioka et al., it is shown that low-temperature conductance measurements allow the detection of fast interface traps with high-frequency signals as small as 1 MHz at 120 K.[28] In this regard, the successful application of the completed High-Low method to WBG systems using routine frequencies at near-cryogenic temperatures appears likely and must be explored. Otherwise, developing higher-frequency (>100 MHz) measurement capabilities is of paramount importance for interface state analysis, even within the context of a completed theory.

## V. Conclusion

In this work, an electrostatic constraint is derived using energy conservation principles to complete the High-Low analysis framework. The derivation recognizes that the dc voltage drop across an MOS structure biased in accumulation is explicitly linked to the voltage drop due to accumulation interface states. In turn, this constraint enables the determination of a physically consistent $C_{OX}$, which is imperative for extracting $D_{IT}$ values in accumulation. The completed framework demonstrated in this work considers the Fermi-Dirac description for $Q_S^+$ to accommodate accumulation conditions, which are likely to violate the bounds of the Maxwell-Boltzmann approximation. However, for systems that exhibit quantum confinement effects, $Q_S^+$ must be calculated by self-consistently solving the Schrödinger and Poisson equations. The validity of the completed High-Low framework is substantiated through simulated high- and low-frequency data generated from an analytical model. The proposed constraint rectifies a fundamental limitation of the High-Low technique, and the completed analysis results in an accurate $D_{IT}$ profile in accumulation for sufficiently fast high-frequency measurements.

## Appendix

### A. A review of the $V_G(\psi_S)$ relationship

During device operation, an externally applied $V_G$ controls the internal $\psi_S$, which dictates the operating modes of the MOS structure. The gate voltage $V_G$ is the sum of voltage drops across the oxide layer, $V_{OX}$, the semiconductor substrate, $V_S$, and the built-in potential across the MOS device due to differences in the metal and semiconductor work functions, $\phi_{MS}$. Therefore, $V_G$ may be described by the following relationship:

$$V_G(\psi_S) = \phi_{MS} + V_{OX}(\psi_S) + V_S(\psi_S). \tag{A1}$$

The voltage drop across the semiconductor $V_S(\psi_S)$ is simply the difference in potential between the semiconductor surface and the semiconductor bulk, or

$$V_S(\psi_S) = \psi_S. \tag{A2}$$

The voltage drop across a dielectric layer is fundamentally described by Gauss's law, which states that the electric field leaving a surface is proportional to the charge enclosed by that surface. Therefore, $V_{OX}(\psi_S)$ must consider the total charge in the semiconductor, $Q_S$, the static charged defects in the oxide, $Q_{OX}$, and the interface state charge, $Q_{IT}(\psi_S)$. The following relationship then describes $V_{OX}$:

$$V_{OX}(\psi_S) = -\frac{1}{C_{OX}}\left(Q_{OX} + Q_{IT}(\psi_S) + Q_S(\psi_S)\right). \tag{A3}$$

The negative sign indicates that a positive charge on the gate corresponds to a negative charge in the underlying layers, and vice versa. The charge contribution of acceptor-like interface states $Q_{IT}$ is a function of $\psi_S$ by the relationship,

$$Q_{IT}(\psi_S) = -q\int_{-\infty}^{\psi_S} D_{IT}(\psi_S)d\psi_S\,, \tag{A4}$$

where the negative sign refers to the respective polarity of the trapped charge. The lower integral limit in (A4) spans to $-\infty$ so that all interface charges with energies below $\psi_S$ that can exchange charge are accounted for. Altogether, (A1) is expanded to be

$$V_G(\psi_S) = \left[\phi_{MS} - \frac{Q_{OX}}{C_{OX}} + \frac{q}{C_{OX}} \int_{-\infty}^{\psi_S} D_{IT}(\psi_S) d\psi_S\right] + \psi_S - \frac{Q_S(\psi_S)}{C_{OX}}. \tag{A5}$$

## B. Discussion on the Fermi-Dirac integral and its calculation

An analytical expression for the semiconductor charge $Q_S$ at the semiconductor surface is derived by solving Poisson's equation while recognizing the appropriate boundary constraints.[8, 10, 13] When Fermi-Dirac statistics are rigorously considered, the derivation involves the Fermi-Dirac integral. Generally, the Fermi-Dirac integral of the $j^{th}$ order is defined as

$$\mathcal{F}_j(\eta_S) = \frac{1}{\Gamma(j+1)} \int_0^{\infty} \frac{\varepsilon^j}{1 + \exp(\varepsilon - \eta_S)} d\varepsilon\,, \tag{B1}$$

where the gamma function, $\Gamma$, extends the factorial function to consider non-integer arguments. Then, $\varepsilon = (E - E_C)/kT$ is the normalized energy over which the improper integral is bounded, and $\eta_S = (E_S - E_C)/kT$ is the normalized surface energy at which the function is evaluated. For an *n*-type structure like the one studied in this work, both $\varepsilon$ and $\eta_S$ reference the conduction band edge, but for a *p*-type structure, these parameters reference the valence band edge, $E_V$. The derivative for $\mathcal{F}_j(\eta_S)$ with respect to $\eta_S$ is given as

$$\frac{d\mathcal{F}_j(\eta_S)}{d\eta_S} = \mathcal{F}_{j-1}(\eta_S). \tag{B2}$$

Extracting $D_{IT}$ using the completed High-Low method may require calculation of $\mathcal{F}_{3/2}\big(\eta_S(\psi_S)\big)$. In general, numerical calculation of $\mathcal{F}_j(\eta_S)$ is completed very simply using common mathematical software (e.g., MATLAB). Alternatively, analytical approximations can provide estimations with a maximum error of ~±0.5%.[29] An approximation for $\mathcal{F}_{3/2}(\eta_S)$ is the following:

$$\mathcal{F}_{3/2}(\eta_S) = \left[\exp(-\eta_S) + 15\sqrt{\pi/2}\left[(\eta_S + 2.64) + (|\eta_S - 2.64|^{2.25} + 14.9)^{4/9}\right]^{-5/2}\right]^{-1}. \tag{B3}$$

Therefore, incorporating $\mathcal{F}_{3/2}(\eta_S(\psi_S))$ in the analysis is computationally straightforward. Then, in the calculation of the degenerate space-charge capacitance $C_{SC}$ discussed in Appendix C, $\mathcal{F}_{1/2}(\eta_S)$ must also be considered, which is similarly approximated as the following:

$$\mathcal{F}_{1/2}(\eta_S) = \left[\exp(-\eta_S) + 3\sqrt{\pi/2}\left[(\eta_S + 2.13) + (|\eta_S - 2.13|^{2.4} + 9.6)^{5/12}\right]^{-3/2}\right]^{-1}. \tag{B4}$$

## C. Analytical expressions for $C_{SC}(\psi_S)$

The space-charge capacitance $C_{SC}(\psi_S)$ is defined as

$$C_{SC(\psi_S)} = \left|\frac{dQ_S(\psi_S)}{d\psi_S}\right|. \tag{C1}$$

For an *n*-type non-degenerate semiconductor, $C_{SC}$ is described as

$$C_{SC}(\psi_S) = \frac{\epsilon_S\epsilon_0}{2L_D}\frac{\left|\exp\left(\frac{q\psi_f}{kT}\right)\left(\exp\left(\frac{q\psi_S}{kT}\right) - 1\right)\right|}{\sqrt{\exp\left(\frac{q\psi_f}{kT}\right)\left(\exp\left(\frac{q\psi_S}{kT}\right) - \frac{q\psi_S}{kT} - 1\right)}}, \tag{C2}$$

where the intrinsic Debye Length $L_D$ is defined as

$$L_D = \sqrt{\frac{\epsilon_S\epsilon_0 kT}{2q^2 n_i}}, \tag{C3}$$

which must not be confused with the extrinsic Debye length $L_{\mathrm{D,Ext.}}$ used in the calculation of $C_{\mathrm{Debye}} = \epsilon_S\epsilon_0/L_{\mathrm{D,Ext.}}$, where $L_{\mathrm{D,Ext.}}$ is defined as

$$L_{\mathrm{D,Ext.}} = \sqrt{\frac{\epsilon_S\epsilon_0 kT}{q^2 N_D}}. \tag{C4}$$

The bulk Fermi potential $\psi_f$ for an *n*-type semiconductor is defined as

$$\psi_f = \frac{kT}{q}\ln\left(\frac{N_D}{n_i}\right). \tag{C5}$$

For an *n*-type degenerate semiconductor that does not consider quantum confinement effects, $C_{SC}$ is found by plugging (14) into (C1) while making use of (B2), which gives

$$C_{SC}(\psi_S) = \frac{\epsilon_S \epsilon_0}{2L_D} \frac{\left| \exp\left(\frac{qE_g}{2kT}\right) \mathcal{F}_{1/2}\left(\eta_s(\psi_S)\right) - \exp\left(\frac{q\psi_f}{kT}\right) \right|}{\sqrt{\exp\left(\frac{qE_g}{2kT}\right) \left[\mathcal{F}_{3/2}\left(\eta_s(\psi_S)\right) - \mathcal{F}_{3/2}(\eta_0)\right] - \exp\left(\frac{q\psi_f}{kT}\right) [\eta_S(\psi_S) - \eta_0]}}. \quad (C6)$$

## D. Simulation parameters

*Table 1: Simulation parameters used in the analytical model based on an n-type 4H-SiC MOS capacitor at room temperature.*

| Parameter | Symbol | Value |
|---|---|---|
| Oxide Capacitance | $C_{OX}$ | $100 \frac{\text{nF}}{\text{cm}^2}$ |
| Semiconductor Bandgap | $E_g$ | 3.23 eV |
| Intrinsic Carrier Concentration | $n_i$ | $4.82 \times 10^{-9} \frac{1}{\text{cm}^3}$ |
| Effective Density of States in the Conduction Band | $N_C$ | $1.63 \times 10^{19} \frac{1}{\text{cm}^3}$ |
| Effective Density of States in the Valence Band | $N_V$ | $2.41 \times 10^{19} \frac{1}{\text{cm}^3}$ |
| Bulk Equilibrium Electron Density | $N_D$ | $1 \times 10^{16} \frac{1}{\text{cm}^3}$ |
| Static Oxide Charge | $Q_{OX}$ | 0 |
| Device Temperature | $T$ | 293 K |
| Semiconductor Dielectric Constant | $\epsilon_S$ | 9.7 |
| Characteristic DC Sweep Residence Time | $\tau_{dc}$ | 1 s |
| Metal-Semiconductor Work Function | $\phi_{MS}$ | 0 |

## Nomenclature

$C_{FB}$ – Flat-band capacitance

$C_{LF}$ – Low-frequency device capacitance

$C_{HF}$ – High-frequency device capacitance

$C_{IT}$ – Interface trap capacitance

$C_{MOS}$ – MOS capacitance measured from a device

$C_{OX}$ – Oxide capacitance

$C_S$ – Semiconductor capacitance

$C_{SC}$ – Space-charge (or depletion) capacitance

$D_{IT}$ – Density of interface traps

$E_C$ – Conduction band-edge energy

$E_g$ – Semiconductor bandgap

$E_f$ – Semiconductor Fermi energy

$E_{FB}$ – Flat-band energy

$E_i$ – Intrinsic semiconductor Fermi energy

$E_T$ – Interface trap energy

$E_V$ – Valence band-edge energy

$f_{ac}$ – AC signal frequency

$f_{OX}$ – $C_{OX}$ correction factor

$\mathcal{F}_j$ – Fermi-Dirac integral of order $j$

$k$ – Boltzmann constant

$L_{\mathrm{D}}$ – Intrinsic Debye length

$n_i$ – Intrinsic carrier concentration

$N_C$ – Effective density of states in the conduction band

$N_V$ – Effective density of states in the valence band

$N_A$ – Bulk equilibrium hole density

$N_D$ – Bulk equilibrium electron density

$Q_{IT}$ – Interface trapped charge

$Q_{OX}$ –Static oxide charge

$Q_S$ – Semiconductor charge

$Q_S^+$ – Semiconductor charge at $\psi_S^+$

$q$ – Fundamental charge constant

$t_{OX}$ – Oxide layer thickness

$T$ – Device temperature

$V_{FB}$ – Flat-band voltage

$V_G$ – Gate voltage

$V_G^+$ – Maximum gate voltage measured in accumulation

$V_{OX}$ – Voltage drop across the oxide

$V_\mathrm{S}$ – Voltage drop across the semiconductor

$\Delta V_{IT}^+$ – Voltage drop due to measured accumulation interface states

$\epsilon_S$ – Semiconductor dielectric constant

$\epsilon_{OX}$ – Oxide dielectric constant

$\epsilon_0$ – Permittivity of free space

$\eta_S$ – Normalized surface Fermi level position

$\eta_0$ – Normalized bulk Fermi level position

$\tau_{dc}$ – Characteristic residence time of the dc sweep

$\tau_{IT}$ – Interface trap time constant

$\phi_{MS}$ – Gate-to-semiconductor work function

$\psi_f$ – Fermi potential of the bulk substrate

$\psi_S$ – Fermi potential at the substrate surface

$\psi_S^+$ – The Berglund constant, or the Fermi potential at the surface corresponding with $V_G^+$

## Acknowledgements

Sandia National Laboratories is a multi-mission laboratory managed and operated by the National Technology and Engineering Solutions of Sandia, LLC, a wholly owned subsidiary of Honeywell International Inc., for the U.S. Department of Energy's National Nuclear Security Administration under Contract DE-NA0003525. This article describes objective technical results and analysis. Any subjective views or opinions that might be expressed in the article do not necessarily represent the views of the U.S. Department of Energy or the U.S. Government.

Sarit Dhar gratefully acknowledges the support provided by the Auburn University Professional Improvement Leave program.

The authors sincerely thank Dr. James Loveless at Sandia National Laboratories for invaluable discussions related to the analytical model.

## Author Declarations

### *Conflicts of Interest*

*The authors have no conflicts to disclose.*

### *Author Contributions*

**Brian D. Rummel**: Conceptualization (lead); Formal Analysis (lead); Investigation (lead); Methodology (lead); Software (lead), Visualization (lead); Writing – Original Draft Preparation (lead); Writing – Review & Editing (lead). **Sarit Dhar:** Validation (equal); Writing/Review & Editing (supporting). **Robert J. Kaplar:** Validation (equal); Writing/Review & Editing (supporting).

## Data Availability

The data that support the findings of this study are available from the corresponding author upon reasonable request.